\documentstyle[12pt]{article}
\begin{document}
\begin{flushright}
VITP 95 - 04\\
December 1995
\end{flushright}
\begin{center}
{\bf RIGHT-HANDED NEUTRINO CURRENTS IN \\
THE $\mbox{SU(3)}_L\otimes \mbox{U(1)}_N$ ELECTROWEAK THEORY}\\
\vspace{2cm}
{\bf Hoang Ngoc Long\footnote{E-mail : hnlong@bohr.ac.vn} }\\
{\it Institute of Theoretical Physics,
National Centre for Natural Science and 
Technology,\\
P.O.Box 429, Bo Ho, Hanoi 10000, Vietnam.}\\
\vspace{1cm}

Abstract\\
\end{center}
A version of the $\mbox{SU(3)}_L\otimes \mbox{U(1)}_N$
electroweak theory in which there are  right-handed neutrino currents
is reconsidered in detail. We argue that in order to have a result
consistent with low-energy one, the  right-handed neutrino component must
be treated as correction instead of an equivalent spin state.
The data from the $Z$-decay allow us to
fix  the limit for $\phi$ as $-0.00285 \leq \phi \leq 0.00018$.
From the neutrino neutral current scattering, we estimate a bound
for the new neutral gauge boson $Z^2$ mass in the range of 400 GeV.
A bound for the  new charged and neutral
(non-Hermitian) gauge bosons $Y^{\pm}, X^o$ is also obtained
from symmetry-breaking hierarchy.

PACS number(s): 12.15.Mm, 12.15.Ff, 12.60.Cn, 12.15.Ji\\
\vspace{3cm}

\begin{center}
Contribution to {\it The Second Rencontres du Vietnam}, October 1995
\end{center}
\newpage
\noindent
{\large\bf I. Introduction}\\[0.3cm]
\hspace*{0.5cm}The standard model (SM)~\cite{gaw} is in great
agreement with present experimental data. Nevertheless, the belief that
some questions remain  unanswered, has resulted
in numerous attempts to find a more fundamental underlying theory.
Therefore many theories beyond the SM  have been proposed~\cite{byee},
and their consequences are now under intensive study.

In the SM, each generation of fermions is
anomaly free. This is true for many extensions of the SM as well,
including the popular grand unified models~\cite{gg}. In these models,
therefore, the number of generations is completely unrestricted on
theoretical grounds.

Recently, an interesting class of alternative models has been
proposed~\cite{svs}
in which each generation is anomalous but different generations are not 
exact replicas of one another, and the anomalies cancel when
a number of generations are taken into account and to be a multiple
of 3. The most economical gauge group which admits  such fermion
representations is $SU(3)_C\otimes
SU(3)_L \otimes U(1)_N$, and it has been proposed by Pisano, Pleitez and
Frampton~\cite{ppf} (for further work on this model see 
Refs.~\cite{fhpp,dng}). The original model did not have right-handed (RH)
neutrinos, but in~\cite{mpp} they have been included. However,
by some reasons, consequences of this model (model I as called in~\cite{mpp})
are too different from those of the SM. For example,
{\it the magnitude of the neutral couplings of the right-handed
neutrinos coincides with that of the left-handed neutrinos
in the SM} (see Eqs.(44,51) in~\cite{mpp}). If so, in order to get
results consistent with low energy phenomenology such as
$\nu_\mu-e$ and $\bar{\nu}_\mu-e$ scatterings, $g_L(e)$ must be
replaced by $g_R(e)$ and vice versa~\cite{bp}. It is obvious that this 
statement is unacceptable.

The purpose of  this paper is to reconsider this model.
It is to be noted that a similar model has been proposed in~\cite{flt}
(for details see~\cite{lt,hnl}) in which RH neutrinos by opposition to
our model, gives contribution to neutral currents
of the left-handed (LH) neutrinos. Since there are many analogies
between the two models in the Higgs and gauge bosons sectors, we will
briefly present them. The main difference arises in neutrino neutral
current scattering, which we will consider in the Section V.2.
By introducing the $Z-Z'$ mixing angle, the exact physical eigenstates
of neutral gauge bosons are obtained. Based on the data from the 
$Z$-decay, the mixing angle is fixed and from the neutrino neutral current
scattering data, we estimate  the $Z^2$ boson mass in the range of 400 GeV,
which is accessible for direct searches at high energy colliders such
as Tevatron, NLC, etc.

This paper is organized as follows. In Sec. II we give a brief review
of the model. The charged and neutral currents are studied in
Sec. III.
In Sec. IV the constraints on the $Z-Z'$ mixing and  masses of the
new gauge bosons  are obtained. Here we argue that in order to have
a result consistent with low energy one, the RH neutrino component
must be treated as correction instead of an equivalent spin state.
Finally, our conclusions are summarized in the last section.\\[0.3cm]
{\large\bf II. The model}\\[0.3cm]
\hspace*{0.5cm}Our model is  based  on the gauge group
\begin{equation}
SU(3)_C\otimes SU(3)_L\otimes U(1)_N.
\label{gg}
\end{equation}
{\it 1. Fermion content and Yukawa interactions}

This  model deals with nine leptons and nine quarks. There are three
left- and right-handed neutrinos ($\nu_e, \nu_\mu ,\nu_\tau$), 
three charged leptons ($e, \mu, \tau$), four quarks with charge 2/3,
and five quarks with charge -1/3.

Under the gauge symmetry (1), the three lepton generations
transform as
\begin{equation}
f^{a}_L = \left( \begin{array}{c}
               \nu^a_L\\ e^a_L\\ (\nu^c_L)^a
               \end{array}  \right) \sim (1, 3, -1/3), e^a_R\sim (1,
1, -1),
\label{l}
\end{equation} 
where a = 1, 2, 3 is the generation index.

Two of the three quark generations transform identically and one generation
(it does not matter which one) 
transforms in a different representation of the gauge group (1):
\begin{equation}
Q_{iL} = \left( \begin{array}{c}
                d_{iL}\\-u_{iL}\\ d'_{iL}\\ 
                \end{array}  \right) \sim (3, \bar{3}, 0),
\label{q}
\end{equation}
\[ u_{iR}\sim (3, 1, 2/3), d_{iR}\sim (3, 1, -1/3), 
d'_{iR}\sim (3, 1, -1/3),\ i=1,2,\]
\[ Q_{3L} = \left( \begin{array}{c}
                 u_{3L}\\ d_{3L}\\ T_{L}
                \end{array}  \right) \sim (3, 3, 1/3),\]
\[ u_{3R}\sim (3, 1, 2/3), d_{3R}\sim (3, 1, -1/3), T_{R}\sim (3, 1, 2/3).\]

Fermion mass generation and symmetry breaking can be 
achieved with just three $SU(3)_{L}$ triplets
\begin{equation}
\chi = \left( \begin{array}{c}
                \chi^o\\ \chi^-\\ \chi^{,o}\\ 
                \end{array}  \right) \sim (1, 3, -1/3),
\rho = \left( \begin{array}{c}
                \rho^+\\ \rho^o\\ \rho^{,+}\\ 
                \end{array}  \right) \sim (1, 3, 2/3),\
\eta = \left( \begin{array}{c}
                \eta^o\\ \eta^-\\ \eta^{,o}\\ 
                \end{array}  \right) \sim (1, 3, -1/3).
\label{h2}
\end{equation}
They are defined by  Yukawa Lagrangians as follows:
\begin{eqnarray}
{\cal L}_{Yuk}^{\chi}&=&\lambda_1\bar{Q}_{3L}T_{R}\chi +
 \lambda_{2ij}\bar{Q}_{iL}d^{'}_{jR}\chi^{*} + \mbox{h.c.}\nonumber\\
 &=&\lambda_1(\bar{u}_{3L}\chi^o+\bar{d}_{3L}\chi^-
+\bar{T}_L\chi^{,o})T_R
+\lambda_{2ij}(\bar{d}_{iL}\chi^{o*}-\bar{u}_{iL}\chi^++
\bar{d}'_{iL}\chi^{,o*})d'_{jR} + \mbox{h.c.}\nonumber\\
{\cal L}_{Yuk}^{\eta}&=&\lambda_{3a}\bar{Q}_{3L}u_{aR}\eta+
\lambda_{4ia}\bar{Q}_{iL}d_{aR}\eta^{*}+\mbox{h.c.}\nonumber\\
&=&\lambda_{3a}(\bar{u}_{3L}\eta^o+\bar{d}_{3L}\eta^-+\bar{T}_L\eta^{,o})
u_{aR}+\lambda_{4ia}(\bar{d}_{iL}\eta^{o*}-\bar{u}_{iL}\eta^+
+\bar{d}'_{iL}\eta^{,o*})d_{aR}+\mbox{h.c.}\nonumber\\
{\cal L}_{Yuk}^{\rho}&=&\lambda_{1a}\bar{Q}_{3L}d_{aR}\rho +
 \lambda_{2ia}\bar{Q}_{iL}u_{aR}\rho^{*}+G'_{ab}\bar{f}^a_Le^b_R\rho+
G_{ab}\varepsilon^{ijk}(\bar{f}^a_{L})_i(f^b_L)^c_j
(\rho^{*})_k+\mbox{h.c.}\nonumber\\
&=&\lambda_{1a}(\bar{u}_{3L}\rho^++\bar{d}_{3L}\rho^o+
\bar{T}_L\rho^{,+})d_{aR}+\lambda_{2ia}(\bar{d}'_{iL}\rho^- 
-\bar{u}_{iL}\rho^{o*}+\bar{d}'_{iL}\rho^{,-})u_{aR}\nonumber\\
& &+ G'_{ab}[\bar{\nu^a_L}\rho^++\bar{e}^a_L\rho^o+(\bar{\nu}^c_L)^a
\rho^{,+}]e^b_R+
G_{ab}[\bar{\nu}^a_L(e^c_L)^b\rho^{,-}
-\bar{e}^a_L(\nu^c_L)^b\rho^{,-}]
+\mbox{h.c.},
\label{yukawa}
\end{eqnarray}
here we have used: $(\bar{\nu}^c_L)^a(\nu^c_L)^b = - \bar{\nu}^b_R\nu^a_R
= 0$,\ $(\nu^c_L)^c=-\nu_L$.

If we have the following vacuum expectation values (VEVs):
$\langle\chi \rangle^T = (0, 0, \omega/\sqrt{2})$,\
$\langle\rho \rangle^T = (0, u/\sqrt{2}, 0)$,\
$\langle\eta \rangle^T = (v/\sqrt{2}, 0, 0)$,
then all fermions gain necessary
masses and the gauge symmetry is broken to the SM gauge symmetry:
\begin{eqnarray}
&\mbox{SU}(3)_{C}&\hspace*{-0.2cm}\otimes \ \mbox{SU}(3)_{L}\otimes 
\mbox{U}(1)_{N}\nonumber \\
&\downarrow      &\hspace*{-0.8cm}\langle \chi \rangle   \nonumber \\
&\mbox{SU}(3)_{C}&\hspace*{-0.2cm}\otimes \ \mbox{SU}(2)_{L}\otimes 
\mbox{U}(1)_{Y}\nonumber \\
&\downarrow      &\hspace*{-0.8cm}\langle \rho \rangle, \langle 
\eta \rangle   \\
&\mbox{SU}(3)_{C}&\hspace*{-0.2cm}\otimes \ \mbox{U}(1)_{Q}.
\nonumber
\label{ssb2}
\end{eqnarray}
Here the electric charge is defined:
$Q=\frac{1}{2}\lambda_3-\frac{1}{2\sqrt{3}}\lambda_8+N$, and
the hypercharge is given:
 $Y=2N-\sqrt{3}\lambda_8/3$ ($\lambda_8=diag(1, 1,
-2)/\sqrt{3}$). In the present model the neutrinos remain massless
at the tree level, and by radiative corrections they will gain 
masses~\cite{wbm}. In this
model, the exotic quarks carry electric charges 2/3 and -1/3,
respectively, similarly to ordinary quarks. Consequently, the exotic
quarks can mix with the ordinary ones. This type of mixing gives the
flavor changing neutral currents (FCNCs).
These FCNCs will be induced due to breakdown of the GIM
mechanism. This type of situation has been discussed previously and
bounds on the mixing strengths can be obtained from the
non-observation of FCNC's in the experiments beyond those
predicted by the SM~\cite{ll}.

{\it 2. Gauge bosons}\\[0.3cm]
\hspace*{0.5cm}As usual, the covariant derivatives are
\begin{equation}
D_\mu  = \partial_\mu  + ig\sum^8_{a=1} W^a_\mu .\frac{\lambda_a}{2}
+ig_N\frac{\lambda^9}{2} N B_\mu ,
\label{derivative}
\end{equation}
where $\lambda^a$(a=1,...,8) are the $SU(3)_L$ generators, and
$\lambda^9=\sqrt{2/3}\  diag(1,1,1)$ is defined such that
$Tr(\lambda^a\lambda^b)=2\delta^{ab}$ and $Tr(\lambda^9\lambda^9)=2$.
Also, $N$ denotes the $N$ charge for the three Higgs multiplets.

The non-Hermitian gauge bosons $\sqrt{2}\ W^+_\mu = W^1_\mu-iW^2_\mu ,
\sqrt{2}\ Y^-_\mu = W^6_\mu- iW^7_\mu ,\sqrt{2}\ X^0_\mu =
W^4_\mu- iW^5_\mu $  have the following masses~\cite{lt,hnl}:
\begin{equation}
M^2_W=\frac{1}{4}g^2(u^2+v^2), M^2_Y=\frac{1}{4}g^2(v^2+\omega^2), 
M^2_X=\frac{1}{4}g^2(u^2+\omega^2).
\label{mnhb}
\end{equation}

The physical neutral gauge bosons are defined through the mixing
angle $\phi$ and $Z,Z'$:
\begin{eqnarray}
Z^1  &=&Z\cos\phi - Z'\sin\phi,\nonumber\\
Z^2  &=&Z\sin\phi + Z'\cos\phi,
\end{eqnarray}
where the photon field $A_\mu$ and $Z,Z'$ are given by~\cite{hnl}:
\begin{eqnarray}
A_\mu  &=& s_W  W_{\mu}^3 + c_W\left(-\frac{t_W}{\sqrt{3}}\ W^8_{\mu}
+\sqrt{1-\frac{t^2_W}{3}}\  B_{\mu}\right),\nonumber\\
Z_\mu  &=& c_W  W^3_{\mu} + s_W\left(-\frac{t_W}{\sqrt{3}}\ W^8_{\mu}+
\sqrt{1-\frac{t_W^2}{3}}\  B_{\mu}\right), \nonumber \\
Z'_\mu &=& \sqrt{1-\frac{t_W^2}{3}}\  W^8_{\mu}+\frac{t_W}{\sqrt{3}}\ B_{\mu}.
\label{apstat}
\end{eqnarray}
The mixing angle $\phi$ is given by
\begin{equation}
\tan^2\phi =\frac{M_{Z}^2-M^2_{Z^1}}{M_{Z^2}^2-M_{Z}^2},
\label{tphi}
\end{equation}
where $M_{Z^1}$ and $M_{Z^2}$ are the {\it physical} mass eigenvalues
\begin{eqnarray}
M^2_{Z^1}&=&\frac{1}{2}\left\{M_{Z'}^2+M_{Z}^2-[(M_{Z'}^2-M_{Z}^2)^2+
4(M_{ZZ'}^2)^2]^{1/2}\right\},\\
M^2_{Z^2}&=&\frac{1}{2}\left\{M_{Z'}^2+M_{Z}^2+[(M_{Z'}^2-M_{Z}^2)^2+
4(M_{ZZ'}^2)^2]^{1/2}\right\},
\end{eqnarray}
with
\begin{eqnarray}
M_{Z}^2   &=&\frac{g^2}{4 c_W^2}(u^2+v^2)=\frac{M_W^2}{c_W^2},\nonumber \\
M_{ZZ'}^2&=&\frac{g^2}{4c_W^2\sqrt{3-4s_W^2}}
\left[u^2-v^2(1-2s_W^2)\right],\\
\label{mzzp} 
M_{Z'}^2 &=&\frac{g^2}{4(3-4s_W^2)}\left[4\omega^2+ \frac{u^2}{c_W^2}
+ \frac{v^2(1-2s_W^2)^2}{c_W^2}\right].
\label{masmat}
\end{eqnarray}

From Eq.(14) we see that $\phi=0$ if $u^2=v^2(1-2s_W^2)$.
Here $W, Z^1$ correspond to the Standard Model charged and neutral
gauge bosons, and there are new gauge bosons $Y^{\pm}, X^o$, and 
$Z^2$.
A precision fit from electroweak observables gives a limit on the
mixing angle (see below) of $-0.00285\leq\phi\leq 0.00018$ and from
 the symmetry-breaking hierarchy $\omega\gg u,v$ Eq.~(\ref{mnhb}) and
Eq.~(\ref{masmat}) lead to
\begin{equation}
M_{Y^+}\simeq M_{X^o}\simeq \frac{\sqrt{3-4s_W^2}}{2}M_{Z^2}
\simeq 0.72 M_{Z^2}.
\label{masx}
\end{equation}
{\large\bf III. Charged and neutral currents}\\[0.3cm]
The interactions among the gauge bosons and fermions are read off directly
from
\begin{eqnarray}
{\cal L}_F & = & \bar{R}i\gamma^\mu(\partial_\mu+ig_NB_\mu N)R\nonumber \\
           &   & + \bar{L}i\gamma^\mu(\partial_\mu+i\frac{g_N}{\sqrt{6}}
B_\mu N + ig\sum^8_{a=1} W^a_\mu . \frac{\lambda_a}{2})L,
\label{current}
\end{eqnarray}
where $R$ represents any right-handed singlet and $L$ any left-handed 
triplet or antitriplet (here, for antitriplets, $\lambda_a$ is
replaced by $-\lambda_a^*$).

The interactions among the charged vector fields with leptons are
\begin{eqnarray}
{\cal L}^{CC}_l& =& - \frac{g}{\sqrt{2}}(\bar{\nu}^a_L\gamma^\mu e^a_LW^+_\mu +
\bar{(\nu^c_L)}^a\gamma^\mu e^a_LY^+_\mu \nonumber \\
& & + \bar{\nu}^a_L\gamma^\mu (\nu^c_L
)^aX^0_\mu + \mbox{h.c.}),
\label{ccl}
\end{eqnarray}
and for the quarks we have
\begin{eqnarray}
{\cal L}^{CC}_q &=&- \frac{g}{\sqrt{2}}[(\bar{u}_{3L}\gamma^\mu d_{3L}+
\bar{u}_{iL}\gamma^\mu d_{iL})W^+_\mu +
(\bar{T}_{L}\gamma^\mu d_{3L}+\bar{u}_{iL}\gamma^\mu d'_{iL})Y^+_\mu 
\nonumber \\
                 & & + (\bar{u}_{3L}\gamma^\mu T_{L}-\bar{d'}_{iL}\gamma^\mu 
d_{iL})X^0_\mu + \mbox{h.c.}].
\label{ccq}
\end{eqnarray}
We can see that the interactions with the $Y^+$ and $X^0$ bosons
violate the lepton number (see Eq.(\ref{ccl})) and the weak isospin 
(see Eq.(\ref{ccq})).

The electromagnetic current for fermions is the usual one
$Q_fe\bar{f}\gamma^\mu fA_\mu$,
and the neutral current interactions can be written in the form
\begin{eqnarray}
{\cal L}^{NC}&=&\frac{g}{2c_W}\left\{\bar{f}\gamma^{\mu} 
[a_{1L}(f)(1-\gamma_5) + a_{1R}(f)(1+\gamma_5)]f 
Z^1_{\mu}\right.\nonumber\\
             & & + \left.\bar{f}\gamma^{\mu} 
[a_{2L}(f)(1-\gamma_5) + a_{2R}(f)(1+\gamma_5)]f Z^2_{\mu}\right\}.
\label{nc}
\end{eqnarray}
The couplings of fermions
with $Z^1$ and $Z^2$ bosons are given as follows:
\begin{eqnarray}
a_{1L,R}(f) &=&\cos\phi\ [T^3(f_{L,R})-s_W^2 Q(f)]\nonumber\\
           & & + c_W^2\left[\frac{3N(f_{L,R})}{(3-4s_W^2)^{1/2}}
-\frac{(3-4s_W^2)^{1/2}}{2c^2_W}Y(f_{L,R})\right]\sin\phi,\nonumber\\
a_{2L,R}(f)&=& - c_W^2\left[\frac{3N(f_{L,R})}{(3-4s_W^2)^{1/2}}
-\frac{(3-4s_W^2)^{1/2}}{2c^2_W}Y(f_{L,R})\right]\cos\phi\nonumber\\
           & & + \sin\phi\ [T^3(f_{L,R})-s_W^2 Q(f)],
\label{vaz}
\end{eqnarray}
where $T^3(f)$ and $Q(f)$ are, respectively, the third component of
the weak isospin and the charge of the fermion $f$. Note that for the exotic
quarks, the weak isospin is equal to zero.
Eqs.~(\ref{vaz}) are valid for both left- and right-handed currents.
Since the value of  $N$ is different for triplets and antitriplets,
the $Z^2$ coupling to left-handed ordinary quarks is different for  the
third family  and thus flavor changing.
Using $\bar{\nu}^c_L\gamma^\mu\nu^c_L=-\bar{\nu}_R
\gamma^\mu\nu_R$ we see that in this model the neutrinos have both
left-handed and right-handed neutral currents:
\begin{eqnarray}
a_{1L}(\nu)&=&\frac{1}{2}\left(\cos\phi +\frac{1-2s_W^2}{
\sqrt{3-4s_W^2}}\sin\phi\right),\
a_{1R}(\nu)=\frac{c^2_W}{\sqrt{3-4s_W^2}}\ \sin\phi,\nonumber\\
a_{2L}(\nu)&=&\frac{1}{2}\left(\sin\phi-
\frac{1-2s_W^2}{
\sqrt{3-4s_W^2}}\cos\phi\right), \
a_{2R}(\nu)=-\frac{c^2_W}{\sqrt{3-4s_W^2}}\ \cos\phi.
\label{van}
\end{eqnarray}

We can also express the neutral current interactions of Eq.~(\ref{nc})
in terms of the vector and axial-vector couplings as follows:
\begin{eqnarray}
{\cal L}^{NC}&=&\frac{g}{2c_W}\left\{\bar{f}\gamma^{\mu} 
[g_{1V}(f)-g_{1A}(f)\gamma_5\right] f Z^1_{\mu}\nonumber\\
             & & + \left.\bar{f}\gamma^{\mu} 
[g_{2V}(f)-g_{2A}(f)\gamma_5]f Z^2_{\mu}\right\}.
\label{ncva}
\end{eqnarray}
The values of these couplings are:
\begin{eqnarray}
g_{1V}(f)&=&\cos\phi\ [T^3(f_L)-2 s_W^2 Q(f)]\nonumber\\
      & & + \sin\phi\left[
\frac{c_W^2}{(3-4s_W^2)^{1/2}}(3N(f_L)+t^2_W N(f_R))-\sqrt{3-4s^2_W}
\frac{Y(f_L)}{2}\right],\nonumber\\
g_{1A}(f)&=&\cos\phi\ T^3(f_L)\nonumber\\
      & & + \sin\phi\left[
\frac{c_W^2}{(3-4s_W^2)^{1/2}}(3N(f_L)-t^2_W N(f_R))-\sqrt{3-4s^2_W}
\frac{Y(f_L)}{2}\right],\nonumber\\
g_{2V}(f)&=&\cos\phi\left[\sqrt{3-4s^2_W}\frac{Y(f_L)}{2}-
\frac{c_W^2}{(3-4s_W^2)^{1/2}}(3N(f_L)+t^2_W N(f_R))\right] \nonumber\\
       & & + \sin\phi\ [T^3(f_L)-2 s_W^2 Q(f)],\nonumber\\
g_{2A}(f)&=&\cos\phi\left[\sqrt{3-4s^2_W}\frac{Y(f_L)}{2}-
\frac{c_W^2}{(3-4s_W^2)^{1/2}}(3N(f_L)-t^2_W N(f_R))\right] \nonumber\\
       & & + \sin\phi\  T^3(f_L).\nonumber
\end{eqnarray}

In the PPF model the coupling strength of $Z^2$ to quarks is much
stronger than that of leptons due to the factor $1/\sqrt{1-4s_W^2}$. 
Therefore, low-energy experiments such as 
neutrino-nucleus scattering and atomic parity violation measurements
would be useful to further constrain the model~\cite{dng}. However,
from (21) it is easy to see
that this  does not happen in this model.

In our model the interactions with the heavy charged and neutral
(non-Hermitian) vector bosons $Y^+, X^o$ violate the lepton
number and the weak isospin. Because of the mixing, the mass eigenstate 
$Z^1$ now picks up flavor-changing couplings proportional to $\sin\phi$.
However, since $Z-Z'$ mixing is constrained to be very small,
evidence of 3-3-1 FCNC's can only be probed indirectly  via
the $Z^2$ couplings.

Whether neutrinos have a right-handed  component is still an
unresolved question. It is possibly the question whose answer will
provide the first evidence of physics beyond the minimal 
SM of particle interactions.
\\[0.3cm]
{\large\bf IV. Constraints on the $Z-Z'$ mixing angle and
the $Z^2$ mass}\\[0.3cm]
\hspace*{0.5cm}There are many ways to get constraints on the 
mixing angle $\phi$ and the $Z^2$ mass. Below we present a
simple one. A constraint on the $Z-Z'$ mixing can be obtained from the
$Z$-decay data, and so we have to calculate the width of the $Z$
in our model.\\[0.3cm]
{\it 1. $Z$ decay modes}\\[0.3cm]
As in the Ref.~\cite{hnl} the total $Z$ width is given by:
\begin{eqnarray}
\Gamma_{total}&=&\Gamma(Z\rightarrow all)=\frac{\bar{\rho_1} G_F}{6\sqrt{2}\pi}
M^3_{Z^1}
\left\{ \cos^2\phi\  \Delta^{SM}_{total}\right.\nonumber\\
              & & + 3\sin2\phi\left[G+\frac{\sqrt{3-4\bar{s}_W^2}}{2} -
\frac{D}{4\sqrt{3-4\bar{s}_W^2}}
 +\delta_{QCD}\left(G-\frac{E}{4\sqrt{3-4\bar{s}_W^2}}\right)\right.\nonumber\\
& & + \left.\left.\frac{\alpha}{12\pi}\left(G-\frac{F}{4\sqrt{3-4\bar{s}_W^2}}
\right)\right] +O(\sin^2\phi)\right\},
\label{ztot}
\end{eqnarray}
where
\begin{eqnarray}
D&=&9-\frac{56}{3}\bar{s}_W^2+\frac{272}{9}\bar{s}_W^4;\
E=3-\frac{20}{3}\bar{s}_W^2+\frac{128}{9}\bar{s}_W^4, \nonumber\\
F&=&33-\frac{332}{3}\bar{s}_W^2+
\frac{1808}{9}\bar{s}_W^4;\
G=\frac{\sqrt{3-4\bar{s}_W^2}}{18}(3-2\bar{s}_W^2)-
\frac{(3-4\bar{s}_W^2)^{3/2}}{36},\nonumber
\end{eqnarray}
and
\[\Delta^{SM}_{total}=\sum_{f=\nu,e,u,d,s,c,b}\{[\bar{g}_V^{SM}(f)]^2+
[\bar{g}_A^{SM}(f)]^2\}(1+\delta_{QED}^f)(1+\delta_{QCD}).\]
Then we get the ratio
\begin{eqnarray}
R^{331}&=&\frac{\Gamma(Z\rightarrow l\bar{l})}{\Gamma_{total}}=R^{SM}_l
\left\{1-\frac{2\tan\phi}{\sqrt{3-4\bar{s}_W^2}}
\left[1+\frac{3\sqrt{3-4\bar{s}_W^2}}{ \Delta^{SM}_{total}}
\right.\right.\nonumber\\
& & \times \left(G+ 
\frac{\sqrt{3-4\bar{s}_W^2}}{2}-\frac{D}{4\sqrt{3-4\bar{s}_W^2}}\right.
+\delta_{QCD}\left(G-\frac{E}{4\sqrt{3-4\bar{s}_W^2}}\right)\nonumber\\
& & + \left.\left.\left.\frac{\alpha}{12\pi}\left(G-\frac{F}{4\sqrt{3-4
\bar{s}_W^2}}\right)\right)\right] +O(\tan^2\phi)\right\},
\label{zrat}
\end{eqnarray}
where $R^{SM}_l$ denotes the SM result: $R^{SM}_l=0.03362$,
for $ \alpha^{-1}(M_Z)=128.87$ ~\cite{rpp,ga},\  $\alpha_s(M_Z)=0.118$,
and for  $\bar{s}_W^2(M_Z)=0.2333$ ~\cite{ll91}.
Taking the experimental result in~\cite{rpp}  $\Gamma=(3.367\pm
0.006)\%$, we obtain bounds for the mixing angle
\begin{equation}
-0.00285 \leq \phi\leq 0.00018.
\label{phi}
\end{equation}

As in Ref.~\cite{hnl} with this mixing angle, $R_b$ in this model still
disagrees with the recent experimental
value $R_b=0.2192\pm 0.0018$ measured at LEP~\cite{blon}.
We hope, however, with the inclusion of new heavy particle loop
effects like exotic quarks, Higgs scalars or of new box diagrams
this result will be improved and consistent with the experimental
data (for recent works on this direction see~\cite{ind}).\\[0.3cm]
{\it 2. Neutrino-electron scattering}\\[0.3cm]
\hspace*{0.5cm}The motivation for focusing on the neutrino neutral current
scatterings is the following: From the theoretical point of view
these reactions
are basic processes free from the complications of strong interactions
and can be used to determine the parameters of the theories. We emphasize
that in the PPF model, these processes are almost the same as in the SM
(for this purpose only neutrino-nucleus scattering and atomic parity 
violation, etc, are suitable). Since in this model neutrinos have both
left- and right-handed current, the effective four-fermion interactions 
relevant to $\nu$-fermion neutral current processes,  are presented 
as follows:
\begin{eqnarray}
-{\cal L}^{\nu f}_{eff}&=&\frac{2 \rho_1 G_F}{\sqrt{2}}\left\{g_{1V}(\nu)
\bar{\nu}\gamma_\mu (1-r\gamma_5)\nu \bar{f}\gamma^\mu[g_{1V}(f)
-g_{1A}(f)\gamma_5]f\right.\nonumber\\
                       & & + \left.\xi g_{2V}(\nu)\bar{\nu}\gamma_\mu
(1-r'\gamma_5)\nu \bar{f}\gamma^\mu[g_{2V}(f)
-g_{2A}(f)\gamma_5]f\right\},
\end{eqnarray}
where  $\xi=\frac{M_{Z_1}^2}{M_{Z_2}^2}$, and 
$r=\frac{g_{1A}(\nu)}{g_{1V}(\nu)},
r'=\frac{g_{2A}(\nu)}{g_{2V}(\nu)} $ are {\it right-handedness} of currents.

The Feynman amplitude for the $\nu_\mu - e$ scattering is
\begin{eqnarray}
T_{if}&=&\frac{2 \rho_1 G_F}{\sqrt{2}}\left\{\bar{\nu}(k')
\gamma_\mu (1-r\gamma_5)\nu (k) \bar{e}(p')\gamma^\mu[g_{1V}(\nu)g_{1V}(e)
-g_{1V}(\nu)g_{1A}(e)\gamma_5]e(p)\right.\nonumber\\
                       & & + \left.\xi\bar{\nu} (k')\gamma_\mu
(1-r'\gamma_5)\nu (k) \bar{e} (p')\gamma^\mu[ g_{2V}(\nu) g_{2V}(e)
- g_{2V}(\nu) g_{2A}(e)\gamma_5]e(p)\right\}.
\end{eqnarray}
In the calculation, it is convenient to introduce  the following symbols:
\begin{eqnarray}
&&L^{\mu\nu}(l,p,p',g_V(l),g_A(l),g'_V(l),g'_A(l))\equiv
Tr\left\{\hat{p}[g_V(l)+g_A(l)\gamma_5]\gamma^{\mu}\hat{p'}
\gamma^{\nu}[g'_V(l)+g'_A(l)\gamma_5]\right\}\nonumber\\
&&=4\left\{[g_V(l)g'_V(l)-g_A(l)g'_A(l)][p^\mu p'^\nu+ p^\nu
p'^\mu-g^{\mu\nu}(p.p')]+i[g_V(l)g'_A(l)-g_A(l)g'_V(l)]\right.\nonumber\\
&&\left.\varepsilon^{\mu\nu\alpha\beta}
p_\alpha p'_\beta\right\},\nonumber\\
&&l_{\mu\nu}(\nu,k,k',r,r')\equiv
Tr\left\{\hat{k}[1+r\gamma_5]\gamma_{\mu}\hat{k'}
\gamma_{\nu}[1+r'\gamma_5]\right\}\nonumber\\
&&=4\left\{[1-r(\nu)r'(\nu)][k_\mu k'_\nu+ k_\nu
k'_\mu-g_{\mu\nu}(k.k')]
+i(r'-r)\varepsilon^{\mu\nu\alpha\beta}
k_\alpha k'_\beta\right\}
\end{eqnarray}
which arise in many trace calculations and have the property
\begin{eqnarray}
&&L^{\mu\nu}(l,p,p',g_{1V}(l),g_{1A}(l),g_{2V}(l),g_{2A}(l))
l_{\mu\nu}(\nu,k,k',r,r')\nonumber\\
&&=32\left\{[g_{1V}(l)g_{2V}(l)-g_{1A}(l)g_{2A}(l)](1-r r')
[(p.k)(p'.k')+(p.k')(p'.k)]\right.\nonumber\\
&&+\left.[g_{1V}(l)g_{2A}(l)-g_{1A}(l)g_{2V}(l)]
(r'- r)[(p.k)(p'.k')-(p.k')(p'.k)]\right\}.
\end{eqnarray}
Taking the square, summing over spin, replacing the spinor products
by projection operators, and neglecting the fermion mass terms, we obtain
\begin{eqnarray}
|T|^2&=&g^2_{1V}(\nu)L^{\mu\nu}(e,p,p',g_{1V}(e),g_{1A}(e),
g_{1V}(e),-g_{1A}(e))l_{\mu\nu}(\nu,k,k',r,-r)\nonumber\\
& & + 2\xi g_{1V}(\nu)g_{2V}(\nu)
L^{\mu\nu}(e,p,p',g_{1V}(e),g_{1A}(e),g_{2V}(e),-g_{2A}(e))
l_{\mu\nu}(\nu,k,k',r,-r')\nonumber\\
      & & + \xi^2g^2_{2V}(\nu)L^{\mu\nu}(e,p,p',g_{2V}(e),g_{2A}(e),
g_{2V}(e),-g_{2A}(e))l_{\mu\nu}(\nu,k,k',r',-r')\nonumber\\
&=&64G^2_F[(p.k)(p'.k')( I^e + J^e ) +(p.k')(p'.k)( I^e - J^e )],
\end{eqnarray}
where
\begin{eqnarray}
I^e&=&g^2_{1V}(\nu)[g_{1V}^2(e)+g_{1A}^2(e)](1+r^2)+2\xi 
g_{1V}(\nu) g_{2V}(\nu)
[g_{1V}(e)g_{2V}(e) + g_{1A}(e)g_{2A}(e)]\nonumber\\
 & & \times (1 + r r') +
\xi^2 g^2_{2V}(\nu)[g^2_{2V}(e) + g^2_{2A}(e)](1+r^{,2}),\nonumber\\
J^e&=&4 rg^2_{1V}(\nu)g_{1V}(e)g_{1A}(e) + 2\xi (r + r')
g_{1V}(\nu)g_{2V}(\nu)[g_{1V}(e)g_{2A}(e)+g_{1A}(e)g_{2V}(e)]\nonumber\\
 & & + 4 \xi^2r'g^2_{2V}(\nu)g_{2V}(e)g_{2A}(e).
\end{eqnarray}
In the laboratory reference frame ($\vec{p}_e =0$), the
cross section is given as in Ref.~\cite{cl}:
\begin{equation}
\frac{d\sigma(\nu_\mu e)}{dE_e}=\frac{1}{32\pi m_e E^2_\nu}
\left(\frac{1}{2.s}\sum |M|^2\right),
\end{equation}
where $E_{\nu},E_e$ are the initial neutrino and final electron energies
and s is the number of the neutrino states. Perform the
usual manipulations~\cite{cl} we get finally
\begin{eqnarray}
\sigma(\nu_\mu e)&=&\frac{\rho^2_1 m_e E_\nu G_F^2}{s\pi}\left[(I^e + J^e)
+\frac{1}{3}(I^e-J^e)
\right],\nonumber\\
\sigma(\bar{\nu}_\mu e)&=&\frac{\rho^2_1 m_e E_\nu G_F^2}{s\pi}
\left[\frac{1}{3}(I^e+J^e)+(I^e-J^e)\right].
\label{gf}
\end{eqnarray}
It is easy to see that for $g_{1V}(\nu)=1/2, r=1$ and
$\xi=0$ we get the SM results.

Substituting coupling constants into Eq.(~\ref{gf}), we finally get
\begin{eqnarray}
\sigma(\nu_\mu e)&=&\frac{\rho^2_1 m_e E_\nu G_F^2}{s 6\pi}\left(
\cos2\phi+\frac{1-2s_W^2}{\sqrt{3-4s_W^2}}\sin2\phi\right)^2\nonumber\\
                 & & \times \left\{(1-4s_W^2+8s_W^4)[(1-\xi)^2+
(r - \xi r')^2]\right.\nonumber\\
                 & & + \left.(1-4s_W^2)(1-\xi)(r - \xi r')
\right\},\\
\sigma(\bar{\nu}_\mu e)&=&\frac{\rho^2_1 m_e E_\nu G_F^2}{s 6\pi}\left(
\cos2\phi+\frac{1-2s_W^2}{\sqrt{3-4s_W^2}}\sin2\phi\right)^2\nonumber\\
                 & & \times \left\{(1-4s_W^2+8s_W^4)[(1-\xi)^2+
(r - \xi r')^2]\right.\nonumber\\
                 & & - \left.(1-4s_W^2)(1-\xi)(r - \xi r')\right\}.
\label{td}
\end{eqnarray}
From Eqs.(35,36), we see that when $\xi=0$, then $\phi = 0,\ r = r'= 1$,
and the low-energy
SM results are obtained if and only if $s=1$. Note that the formulas
(35,36) are valid for
theories with one extra neutral gauge boson $Z^2$ and neutrinos having
left- and right-handed components.

For this model we have:
\begin{eqnarray}
r(\nu)&\simeq&1-\frac{4c_W^2}{\sqrt{3-4s_W^2}}\tan\phi + O(\tan^2\phi),
\nonumber\\
r'(\nu)&\simeq&-\frac{1}{3-4s_W^2}\left(1+
\frac{4c^2_W}{\sqrt{3-4s_W^2}}\tan\phi\right) + O(\tan^2\phi).
\label{rrph}
\end{eqnarray}

As in~\cite{hnl}, taking an average value for $\phi=-0.00134$,  $m_t=174$ GeV
~\cite{ga}, the running $s_W^2=0.21$ and the experimental
results on $\frac{\sigma(\bar{\nu}_\mu e)}{E_\nu}=(1.17\pm 0.206)
\times 10^{-42} cm^2/GeV$ and
$\frac{\sigma(\nu_{\mu}e)}{E_{\nu}}=(1.8\pm 0.32)\times 10^{-42}
cm^2/GeV$  given
in~\cite{ah} the allowed range of the new gauge boson masses
are $M_{Z^2} \geq 320$ GeV and 280 GeV , respectively . Thus
Eq.~(\ref{masx})
gives a limit for the masses of the gauge bosons $Y^{\pm}, X^o$:
$ M_X  \geq 230$ GeV and 200 GeV, respectively.

In our model, the free parameters are $\sin^2\theta_W, M_{Z^2}$, and
$\phi$
which are constrained from experiment. $M_{Z^1}$ is related by Eq.~(\ref{tphi}) where $M_Z=
M_W/\cos\theta_W$ is the prediction for the $Z$ mass in the absence of 
mixing $\phi=0$. It is interesting to consider the special case $\phi=0$.
We then have $M_{Z^1}=M_Z, \rho_1=1$ and $s_W(M_{Z^1})=s_W(M_Z)$.
From Eq.(35) and Eq.(36) we get bounds for the new gauge bosons $Z^2$ mass:
$M_{Z^2}\geq 460$ GeV and 350 GeV, respectively.
Thus, the only way to get a rigorous bound of $M_{Z^2}$
is through  low energy processes as
considered in this paper. Our bounds could be improved
significantly with more precise data.\\[0.3cm]
{\large\bf V. Discussion}\\[0.3cm]
\hspace*{0.5cm}In this paper, we reconsidered and presented
a further development of the 331 model with neutrino right-handed currents.
We have shown that this model has some advantages over the original
331 model. First, in the Higgs sector, we need only three Higgs
triplets for generating fermions and gauge bosons  masses as
well as for breaking the gauge symmetry.
Moreover in the limit $\phi = 0$, all couplings of the
ordinary fermions to $Z^1$ boson are the same as in the SM.
In this model there is no limit for the Weinberg angle $\sin^2\theta_W
< \frac{3}{4}$.

The lepton number is violated  in 
the heavy charged and neutral (non-Hermitian) vector bosons
interactions.
We also have flavor-changing neutral currents in the quark sector
coupled to the new $Z^2$ boson. All the heavy bosons have  masses
depending on  $\langle\chi\rangle$ and this VEV is, in principle, arbitrary.
We argue that in order to get the low energy SM result right-handed
component of neutrinos in this model has to
be considered as a correction instead of an equivalent spin state
(spin-average factors of $\frac{1}{2}$).

Finally, we emphasize
again that  experimental data from the $Z$-decay and
$\bar{\nu}_{\mu}- e ,\  \nu_{\mu}-e$ scattering processes allows us to
estimate the mixing angle $\phi$ and the new gauge boson masses.
To get stronger limits we have to consider other parameters such as
the left-right cross section  asymmetry, $N_\nu$, etc.

To summarize, we have shown that because of the $Z-Z'$ mixing, there is
a modification to the $Z^1$ coupling proportional to $\sin\phi$,
and the $Z$-decay gives  $-0.00285\leq \phi \leq 0.00018$. The data from
neutrino neutral current elastic scatterings shows that mass of the
new neutral gauge boson  $M_{Z^2}$ is in the range of 400 GeV (90\%
C.L.), and 
from the symmetry-breaking hierarchy we get:
$M_{Y^+}\simeq M_{X^o}\simeq 0.72 M_{Z^2} \geq 290$ GeV.
It may well be that the next stage of the developments will consist
of the discoveries of more sequential right-handed neutrinos.
\\[0.3cm]

{\large\bf Acknowledgement}\\[0.4cm]
\hspace*{0.5cm}It is a pleasure to acknowledge Dr.
R. Foot and  Prof. C. Verzegnassi  for useful  discussions.
I would  like to thank Prof. Abdus Salam, Prof. S. Randjbar-Daemi,
Prof. J. Strathdee, the International
Atomic Energy Agency and UNESCO for hospitality at the
International Centre for Theoretical Physics, Trieste.
I thank Prof. P. Aurenche and Prof. X. Y. Pham for kind 
support and help.\\[0.5cm]


\begin{thebibliography}{99}
\bibitem{gaw}S. L. Glashow, Nucl. Phys. {\bf 20}, 579 (1961); A. Salam,
in Elementary Particle Theory, ed. N.Svartholm, (1968);
S. Weinberg, Phys. Rev. Lett. {\bf 19}, 1264 (1967).
\bibitem{byee} G. G. Ross, {\it Grand Unified Theories} (Cambridge
University Press. Cambridge, 1987), and references therein.
\bibitem{gg}H. Georgi and S. L. Glashow, Phys. Rev. Lett.
{\bf 32}, 438 (1974). For a review, see P. W. Langacker, Phys. Rep.
{\bf C72}, 185 (1981).
\bibitem{svs}M. Singer, J. W. F. Valle, and J. Schechter, Phys. Rev.
{\bf D22}, 738 (1980).
\bibitem{ppf} F. Pisano and V. Pleitez, Phys. Rev. {\bf D46}, 410 (1992);
P. H. Frampton, Phys. Rev. Lett. {\bf 69}, 2889 (1992).
\bibitem{fhpp} R. Foot, 
O.F. Hernandez, F. Pisano, and V. Pleitez, Phys. Rev. {\bf D47}, 4158 (1993).
\bibitem{dng}Daniel Ng, Phys. Rev. {\bf D49}, 4805 (1994).
\bibitem{mpp} J. C. Montero, F. Pisano, and V. Pleitez, Phys. Rev.
{\bf D47}, 2918 (1993).
\bibitem{wbm}L. Wolfenstein, Nucl. Phys. {\bf B186}, (1981) 147;
R. Barbieri and R. N. Mohapatra, Phys. Lett. {\bf B218}, (1989) 225.
\bibitem{bp}V. Barger and R. Phillips, {\it Collider Physics},
Frontiers in Physics 71, Addison-Wesley (1987), Chapters 3 \& 4.
\bibitem{flt} R. Foot, H. N. Long, and Tuan A. Tran, 
Phys. Rev. {\bf D50}, R34 (1994). 
\bibitem{lt} H. N. Long and T. A. Tran, Mod. Phys. Lett. {\bf A9}
(1994) 2507.
\bibitem{hnl} H. N. Long, [hep-ph/9504274], Phys. Rev. {\bf D53},
437 (1996).
\bibitem{ll}See, for example, P. W. Langacker and D. London, Phys. Rev.
{\bf D38}, 886 (1988).
\bibitem{rpp} Particle Data Group, L. Montanet {\it et al}., Phys. Rev.
{\bf D50}, 1357 (1994).
\bibitem{blon} A. Blondel, in {\it Physics at LEP 200 and Beyond}
Proceedings of the Workshop on Elementary Particle Physics, Teupitz,
Germany, 1994, edited by T. Riemann and J. Blumlein [Nucl. Phys. B
(Proc. Suppl.) {\bf 37B} (1994)].
\bibitem{ll91}P. Langacker and M. Luo, Phys. Rev. {\bf D44}, 817 (1991).
\bibitem{bv}A. Blondel and C. Verzegnassi, Phys. Lett. {\bf B311} (1993)
346.
\bibitem{ga} G. Altarelli, CERN preprint CERN-TH/95-3 (1995).
\bibitem{pll} P. Langacker and M. Luo, Phys. Rev. {\bf D45}, 278 (1992);
J. Hewett and T. Rizzo, Phys. Rep. {\bf 183}, 193 (1989);
Phys. Rev. {\bf D45}, 161 (1992), and references therein.
\bibitem{ind} Xu Wang, J. L. Lopez, and D. V. Nanopoulos, 
Phys. Rev. {\bf D52},4116 (1995) ; D. Ng, TRI-PP-95-8; 
G. Bhattacharyya, D. Choudhury,
and K. Sridhar, Phys. Lett. {\bf B355}, (1995) 193; J. Ellis,
J.L. Lopez, and D.V. Nanopoulos, preprint CERN - TH/95-314.
\bibitem{cl}T. P. Cheng and L. F. Li, {\it Gauge Theory of
Elementary Particle Physics}, ( Oxford University Press, NY 1984).
\bibitem{ah}L. A. Ahrens {\it et al}.,\ Phys. Rev. {\bf D41}, 3297 (1990).
\end{thebibliography}
\end{document}